\documentclass[12pt]{article}
\usepackage{fullpage}
\usepackage{amsmath}
\usepackage{amssymb}
\usepackage{amsthm}
\usepackage{graphicx}
\usepackage{fancyhdr}
\usepackage{feynmf}
\usepackage{cite}
\usepackage{hyperref}

\usepackage{color}
\definecolor{darkblue}{rgb}{0,0,0.5}

\begin{document}

\title{{\bf An Unorthodox Introduction to QCD}}
\author{Andrew Larkoski\footnote{larkoski@reed.edu}\\ \\\small{Physics Department}\\\small{Reed College}}
\date{\today}
\maketitle

\begin{abstract}
These are lecture notes presented at the 2017 CTEQ Summer School at the University of Pittsburgh and the 2018 CTEQ Summer School at the University of Puerto Rico, Mayag\"uez.  The title is a reference to \cite{Strassler:2003qg} and introduces perturbative QCD and its application to jet substructure from a bottom-up perspective based on the approximation of QCD as a weakly-coupled, conformal field theory.  Using this approach, a simple derivation of the Sudakov form factor with soft gluon emission modeled as a Poisson process is presented.  Topics of the identification and discrimination of quark- versus gluon-initiated jets and jet grooming are also discussed.
\end{abstract}

\clearpage

\section{Lecture 1: Consequences of Scale Invariance in QCD}

It is extremely challenging to introduce any subject in two hours, and the field of jet substructure is especially challenging because its purview now encompasses much or most of the physics program of the Large Hadron Collider (Higgs, beyond the Standard Model, Standard Model measurements, fragmentation, heavy flavor, heavy ions, etc. See \cite{Larkoski:2017jix,Adams:2015hiv,Altheimer:2013yza,Altheimer:2012mn,Abdesselam:2010pt} for reviews.).  So, these lectures will be narrow in scope and ignore much or most of the applications of this extremely exciting field.  While you have probably been introduced to QCD from the gauge principle, analogies with electromagnetism, and finally its (high-energy) Lagrangian, I want to take a different approach here.  Jet substructure, at its most fundamental, is a study of QCD in the near-soft (low-energy) and collinear (small-angle) limits, and the Lagrangian of QCD isn't a natural starting point for studying this.  Instead of a ``top-down'' approach, I want to emphasize a ``bottom-up'' approach, starting from some natural, simple assumptions about the behavior of QCD at high energies.  We'll see that this will be remarkably powerful.

To proceed, we need to make two reasonable assumptions.  These are important enough that I will call them axioms, for the purposes of this lecture.
\begin{quote}
\noindent \underline{Axiom 1:} At high energies, the coupling of QCD, $\alpha_s$, is small.  Therefore, quarks and gluons are good quasi-particles.
\end{quote}
This axiom essentially means that the perturbation theory of QCD (defined by Feynman diagrams, for example) is a good approximation.  It is sensible to describe final states in terms of quarks and gluons as corrections to this picture are small because $\alpha_s$ is small.
\begin{quote}
\noindent \underline{Axiom 2:} At high energies, QCD has no intrinsic scales.  Quarks and gluons are massless, and so QCD is (approximately) a conformal, or scale-invariant, quantum field theory.
\end{quote}
We know that this axiom strictly isn't true.  While quarks and gluons may be very low mass or massless, hadrons are massive.  Also, the coupling $\alpha_s$ runs with energy, spoiling true scale invariance.  Nevertheless, because $\alpha_s$ is small (by axiom 1), these deviations from scale invariance can be though of as corrections.  This is how we will treat them in this lecture, and I will discuss how to fix-up this picture later.

With these axioms established, I would like to do some simple calculations.  Let's calculate the probability for a quark to emit a gluon:
\begin{equation}
P(q\to qg) = \left|\,\raisebox{-0.42\height}{\includegraphics[width=5cm]{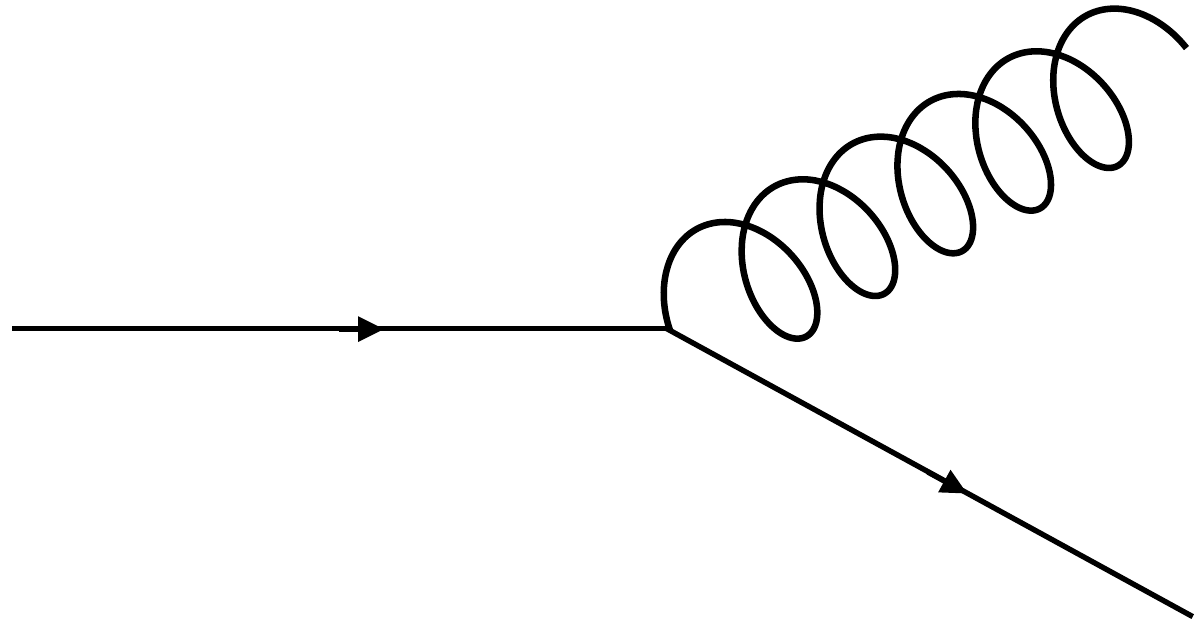}}\,\right|^2\,.
\end{equation}
We will express this probability in terms of the phase space variables of the final state (the quark and the gluon).  What are these phase space variables?  For two particles, two-body phase space is two-dimensional.  Each particle has a four-vector momentum which is required to be on-shell and massless.  Additionally, the sum of those four-vector momenta is the total initial momentum.  This leaves two degrees of freedom, or two phase space variables.  We will choose these phase space variables to be the energy of the gluon, $E_g$, and the invariant mass of the final quark and gluon, $m^2$.  Then,
\begin{equation}
P(E_g,m^2) = \left|\,\raisebox{-0.35\height}{\includegraphics[width=5cm]{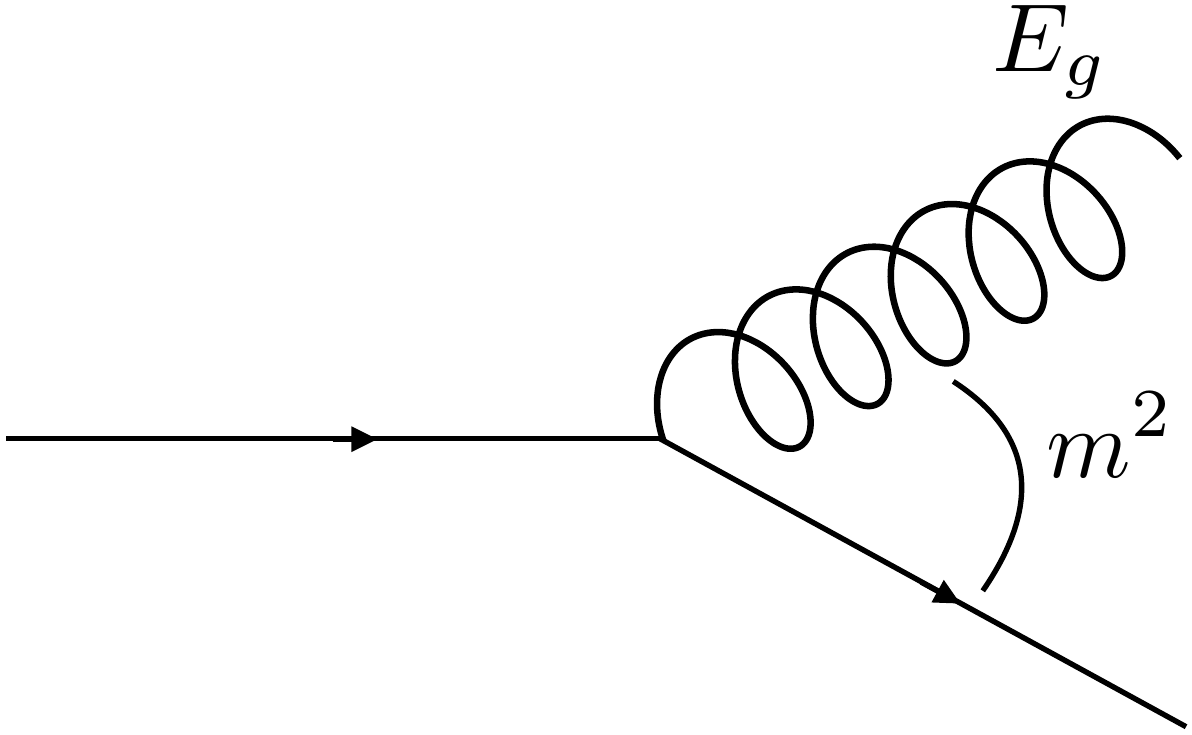}}\,\right|^2\,.
\end{equation}
Note that $m^2 = 2p_q\cdot p_g = 2E_qE_g(1-\cos\theta_{qg})$.

What can this probability be?  Our assumption of scale-invariance helps us out.  Scale invariance means that the probability is unchanged if the energy or mass scales are multiplied by a factor $\lambda > 0$:
\begin{equation}
P(\lambda E_g,\lambda^2m^2)\, d(\lambda E_g)\, d(\lambda^2 m^2) = P(E_g,m^2)\, dE_g\, dm^2\,.
\end{equation}
What could this function be?  The simplest function that one can write down is
\begin{equation}
P(E_g,m^2)\, dE_g\, dm^2 = \frac{\alpha_s C_F}{\pi}\frac{dE_g}{E_g}\frac{dm^2}{m^2}\,.
\end{equation}
Before continuing, I should say a couple things about this expression.  First, the overall factor of $\alpha_s C_F / \pi$ is the strength to which a gluon couples to a quark; $C_F$ is the color factor that represents the amount of color that the quark carries (called the fundamental representation Casimir).  We'll come back to this later.  Note also that we could multiply this expression by any function of $E_g^2/m^2$ and still maintain scale-invariance.  This will be important for detailed studies, but there is a well-defined approximation in which we can ignore such terms.  This is called the ``double-logarithmic approximation'' or DLA.

With this DLA probability in hand, let's change variables to dimensionless quantities, as they are a bit nicer to work with.  Let's express the probability in terms of the gluon's energy fraction, $z$, and the angle $\theta_{qg}\equiv \theta$ between the quark and the gluon:
\begin{align}
&z = \frac{E_g}{E_q+E_g}\,, &1-\cos\theta = \frac{m^2}{2E_qE_g}\,.
\end{align}
Then, the probability becomes
\begin{equation}
P(z,\cos\theta) \, dz\, d\cos\theta= \frac{\alpha_s C_F}{\pi}\frac{dz}{z}\frac{d\cos\theta}{1-\cos\theta}\,.
\end{equation}
Let's even go one step further and work in the small angle limit, $\theta\ll 1$.  Then,
\begin{equation}
P(z,\theta^2)\, dz \, d\theta^2 \to \frac{\alpha_s C_F}{\pi}\frac{dz}{z}\frac{d\theta^2}{\theta^2}\,.
\end{equation}
This expression tells us a huge amount of physics.  Note that the probability diverges when either $z\to 0$ or $\theta\to 0$, in the soft and/or collinear limits. It seems weird for a probability to diverge, but we just have to reinterpret it.

Consider, for example, the soft limit, $z\to 0$.  If the energy of the gluon $E_g\to 0$, then what distinguishes that final state from just the quark, with no gluon?
\begin{center}
\raisebox{-0.6\height}{\includegraphics[width=3.5cm]{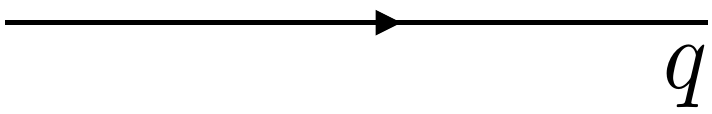}} \qquad vs.\qquad \raisebox{-0.15\height}{\includegraphics[width=3.5cm]{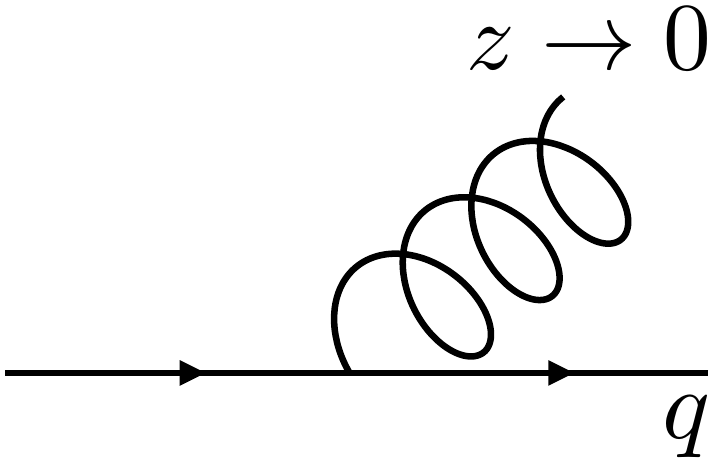}}\qquad?
\end{center}
Is there a measurement we can do to distinguish these systems?  The answer is no!  They become degenerate in the $z\to 0$ limit.  Indeed, Feynman diagram perturbation theory is degenerate perturbation theory, which is why the probability diverges in the $z\to 0$ limit.  There is no measurement we can do to distinguish a system with no gluons, one 0 energy gluon, two 0 energy gluons, three 0 energy gluons, etc.  Results and predictions in degenerate perturbation theory are only finite when we sum up all degenerate states as guaranteed by the Kinoshita-Lee-Nauenberg theorem \cite{Kinoshita:1962ur,Lee:1964is}.  We will see how to do this in a second.  As $z\to 0$, we should not interpret $P(z,\theta^2) \, dz \, d\theta^2$ as a probability, but rather as an expectation value of the number of soft/low energy gluons emitted from the quark.

Similar arguments follow for the collinear limit, $\theta^2\to 0$, but I won't discuss that in detail.

Let's rewrite the probability in an enlightening way:
\begin{equation}
P(z,\theta^2)\, dz \, d\theta^2 = \frac{\alpha_s C_F}{\pi}\frac{dz}{z}\frac{d\theta^2}{\theta^2}= \frac{\alpha_s C_F}{\pi}\, d\left(\log \frac{1}{z}\right)\, d\left(\log\frac{1}{\theta^2}\right)\,.
\end{equation}
That is, emissions of soft/collinear gluons are uniformly distributed in the $(\log 1/z, \log1/\theta^2)$ plane!  There's a very nice way to visualize this, in what is called a ``Lund diagram'' \cite{Andersson:1988gp}.  This is:
\begin{center}
\includegraphics[width=9cm]{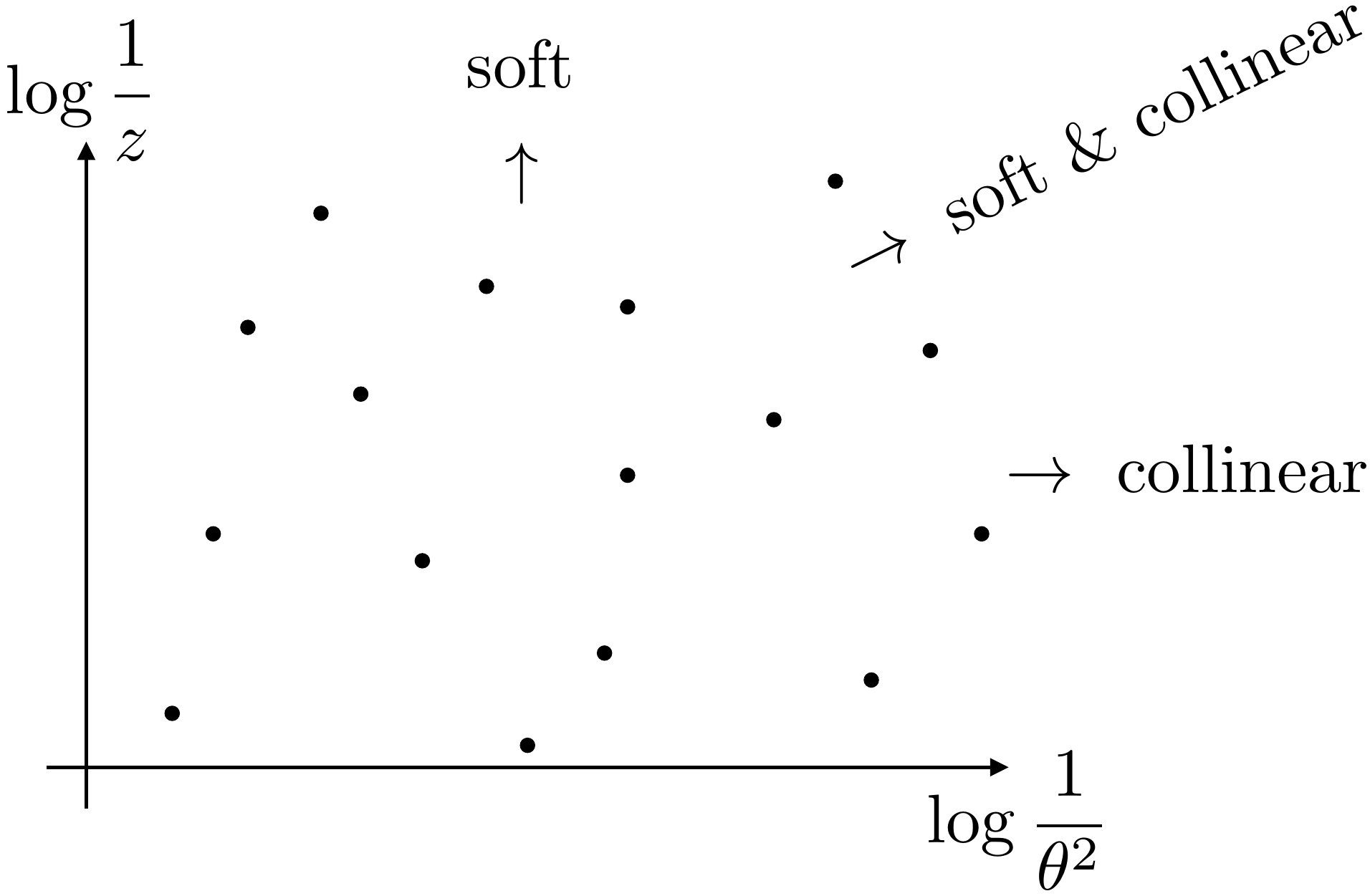}
\end{center}
Here, each $\bullet$ denotes another gluon emission off of the quark, and the emissions are uniformly distributed in the plane.  This is a semi-infinite plane and depending on how we approach $\infty$, we are sensitive to a different singular limit.  Moving vertically in the plane is the soft limit, horizontally is the collinear limit, and diagonally is the soft and collinear limit.

At this point, I should emphasize that this uniform distribution of emissions is special to our approximations.  Including a running coupling, higher-order effects, hadronization (which cuts off this picture at some point), etc., will change this picture.  Nevertheless, there is a sense in which all of those things are corrections to this simple picture.  Additionally, filling out this plane is the goal of Monte Carlo parton shower programs, like {\sc Pythia} \cite{Sjostrand:2006za,Sjostrand:2014zea} and {\sc Herwig} \cite{Bahr:2008pv,Bellm:2015jjp}.  They each employ different methods for doing so, but their fundamental goal is the same.

We could stop here, but I want to do a now-trivial calculation since we have set up this framework.  Let's calculate the distribution of the ratio of the invariant mass of the quark-gluons system to its total energy:
\begin{equation}
\tau = \frac{m^2}{E^2}\,.
\end{equation}
In our phase space coordinates and with our assumptions, the observable $\tau$ is:
\begin{equation}
\tau = \sum_{i=\text{gluon}}z_i \theta_i^2\,,
\end{equation}
where $z_i$ is the energy fraction of the $i^{\text{th}}$ gluon and $\theta_i$ is the angle of the $i^{\text{th}}$ gluon to the quark.  (I use the symbol $\tau$ for this observable because it is identical to thrust \cite{Farhi:1977sg} in the soft and/or collinear limits.)  The sum runs over all emitted gluons/$\bullet$ emissions in the $(\log 1/z, \log1/\theta^2)$ plane.  We will calculate the cumulative probability distribution, $P(x < \tau)$; that is, the probability the measured value of this observable is less than some value $\tau$.

To do this, note that the emissions are uniformly distributed in $(\log 1/z, \log1/\theta^2)$.  This means that in ``real'' space $(z, \theta^2)$, emissions are exponentially far apart!  This will help dramatically simplify our task.  Because of this observation, there is a single emission that dominates the value of $\tau$, and all others provided tiny corrections.  So, with emissions in the plane as:
\begin{center}
\includegraphics[width=7.5cm]{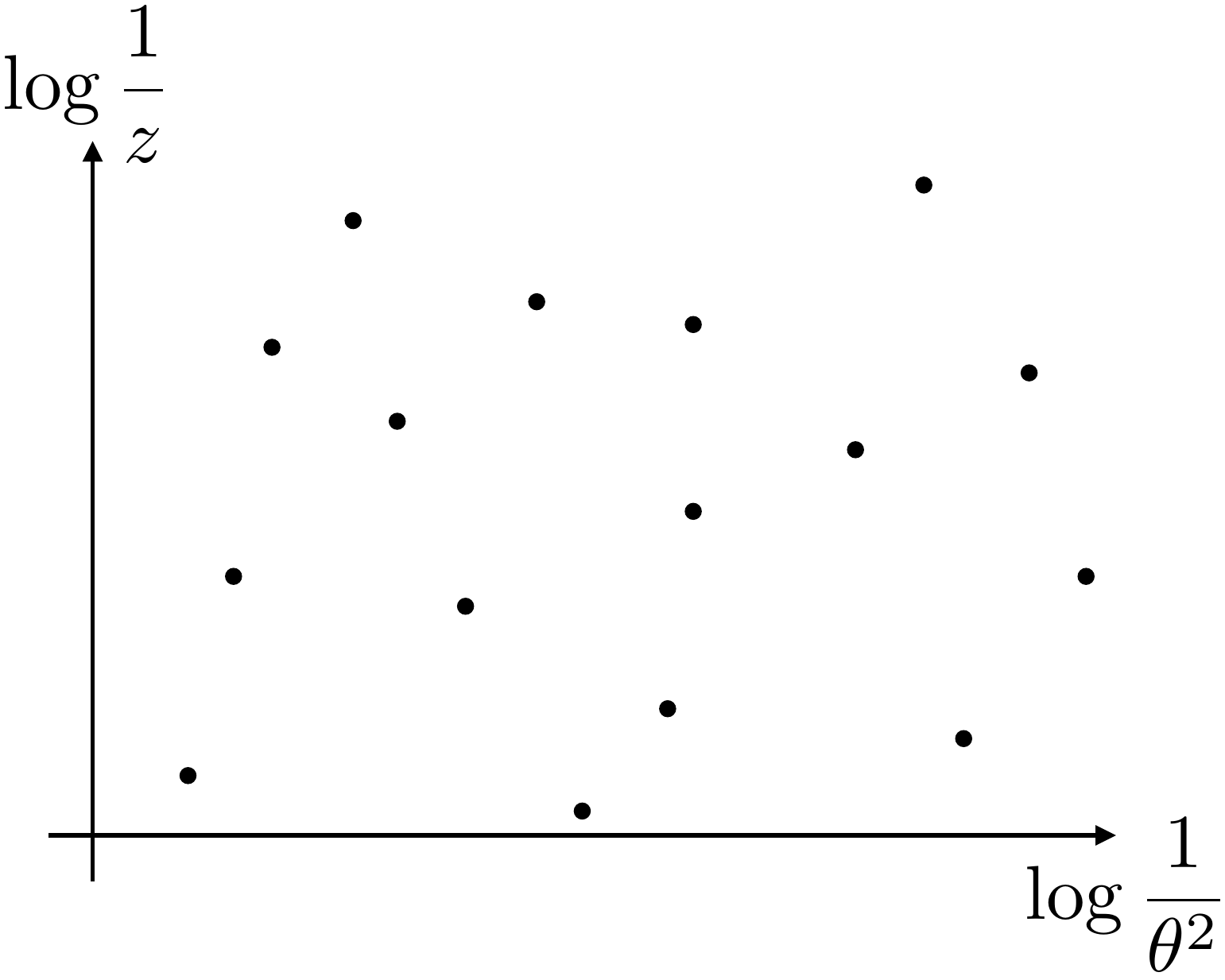}
\end{center}
there will be one that dominates the value of $\tau$: $\tau = z\theta^2$.  Note that a fixed value of $\tau$ on this plane corresponds to a line:
\begin{equation}
\log \tau = \log z+\log\theta^2\,.
\end{equation}
This line then corresponds to
\begin{center}
\includegraphics[width=8.5cm]{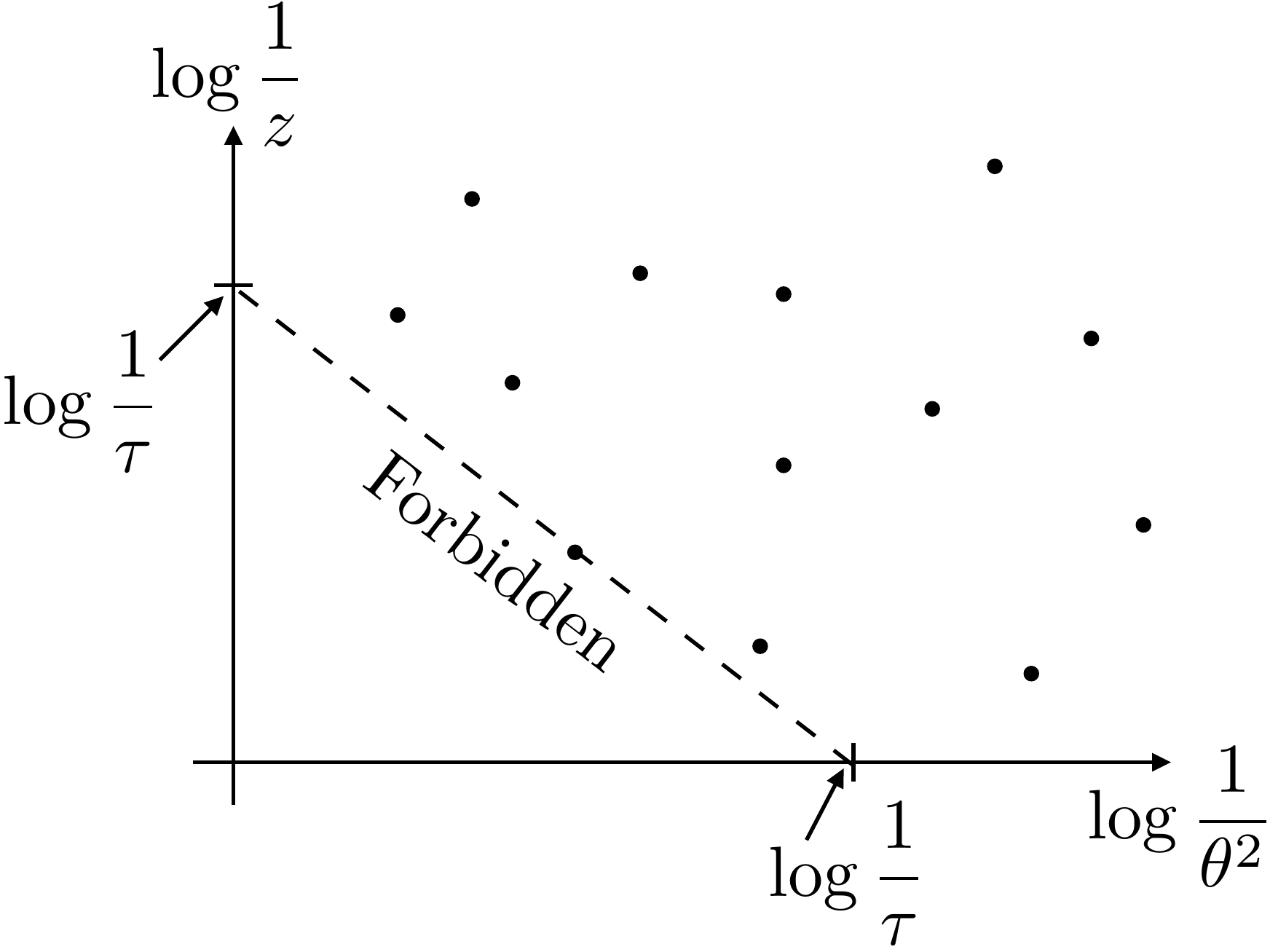}
\end{center}
All emissions above the line are tiny corrections, there is one emission on the line, and no emissions below the line.  If there were emissions below the line, then the measured value of $\tau$ would have increased.  So, for calculating the cumulative probability, we must calculate the probability that there were no emissions below the line.

This probability is easy to calculate.  We can imagine breaking up the forbidden triangle into many regions:
\begin{center}
\includegraphics[width=4cm]{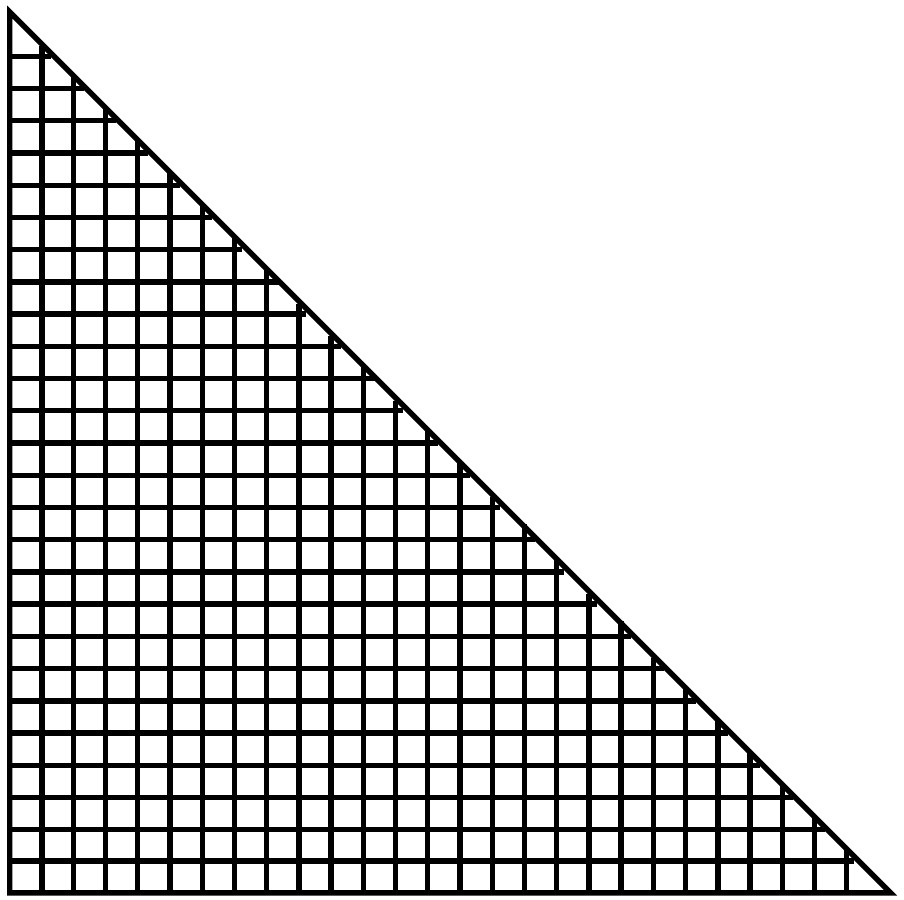}
\end{center}
The probability for emission into any one region is proportional to the area of the region:
\begin{equation}
P(\text{emit in region }i) = \frac{\alpha_s C_F}{\pi}\cdot (\text{Area of region }i)\,.
\end{equation}
Therefore, the probability of no emissions is 1 minus this:
\begin{equation}
P(\text{no emit in region }i) = 1-\frac{\alpha_s C_F}{\pi}\cdot (\text{Area of region }i)\,.
\end{equation}
If we break up the forbidden triangle into $N$ equal-area regions then the area of any one region is
\begin{equation}
\text{Area of region }i = \frac{\frac{1}{2}\log^2\tau}{N}\,,
\end{equation}
because the area of the triangle is $\frac{1}{2}\log^2\tau$.  Then, to forbid any emission in all regions, we multiply these probabilities together:
\begin{equation}
P(\text{no emissions}) = \left(
1-\frac{\frac{\alpha_s}{\pi}\frac{C_F}{2}\log^2\tau}{N}
\right)^N\,.
\end{equation}
Taking the limit as $N\to\infty$, this transmogrifies into an exponential:
\begin{equation}
P(\text{no emissions}) = \exp\left[
-\frac{\alpha_s}{\pi}\frac{C_F}{2}\log^2\tau
\right]\,.
\end{equation}
This is just equal to the cumulative probability
\begin{equation}
P(x< \tau)=\exp\left[
-\frac{\alpha_s}{\pi}\frac{C_F}{2}\log^2\tau
\right]\,.
\end{equation}
Note that this is exponentially suppressed as $\tau\to 0$.  This object is called the Sudakov form factor \cite{Sudakov:1954sw}.

To find the probability distribution, we just differentiate:
\begin{equation}
p(\tau) = \frac{d}{d\tau}\exp\left[
-\frac{\alpha_s}{\pi}\frac{C_F}{2}\log^2\tau
\right] = -\frac{\alpha_s C_F}{\pi}\frac{\log\tau }{\tau}\exp\left[
-\frac{\alpha_s}{\pi}\frac{C_F}{2}\log^2\tau
\right]\,.
\end{equation}
We've tamed all the infinities!  The Sudakov form factor is an explicit sum over all degenerate states with soft/collinear gluon emission.  The probability distribution is finite, and in fact 0 for $\tau\to 0$.

Before concluding this lecture, I want to connect this to a fundamental problem in jet physics: discrimination of quark-initiated jets from gluon-initiated jets.  We can perform the same exercise for gluon jets, and we find the cumulative distribution:
\begin{equation}
P_g(x< \tau)=\exp\left[
-\frac{\alpha_s}{\pi}\frac{C_A}{2}\log^2\tau
\right]\,.
\end{equation}
The only change is replacing $C_F$ by $C_A$, which is the color Casimir for the adjoint representation (the color carried by the gluon).  Schematically, the distributions of $\tau$ for the quark and gluon jets look like:
\begin{center}
\includegraphics[width=6.5cm]{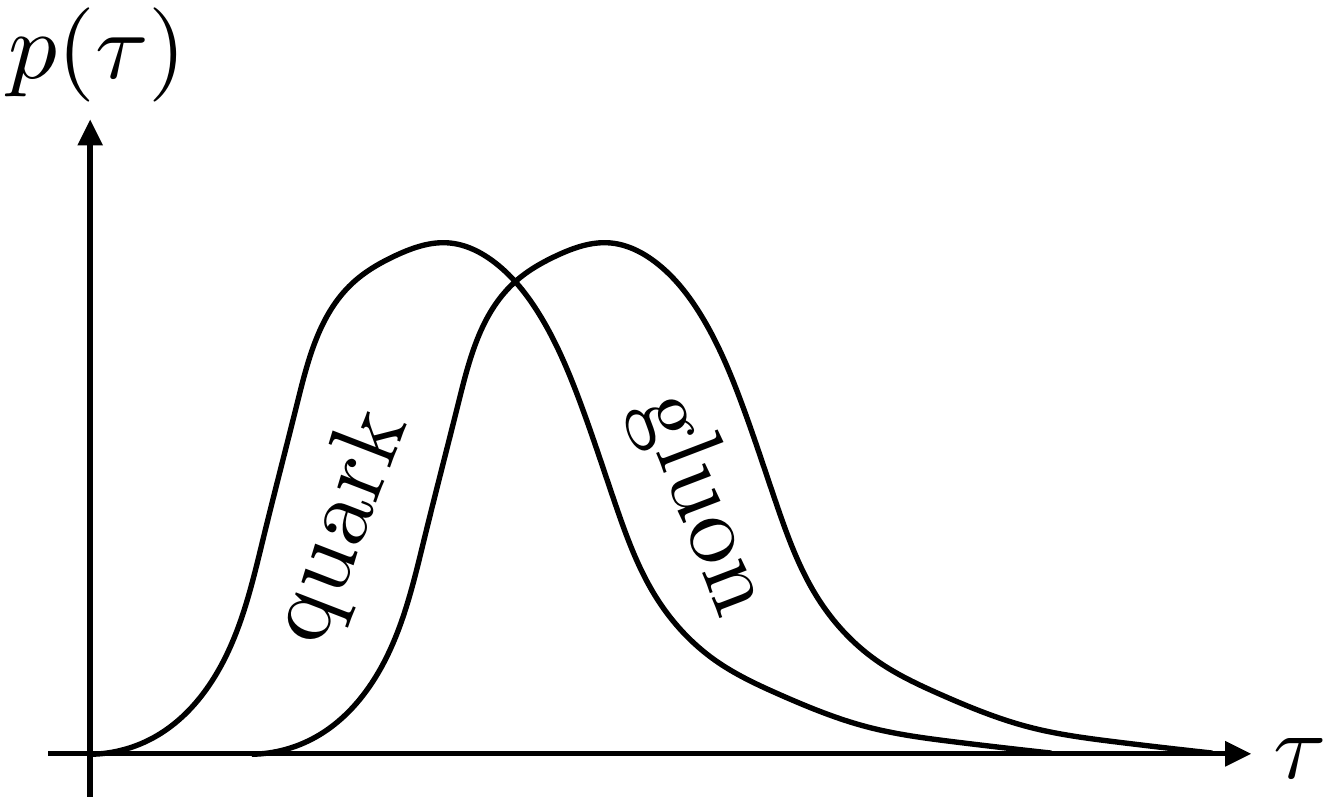}
\end{center}
The ratio between the average values of these distributions is controlled by the ratio of $C_A$ to $C_F$.

To separate quarks from gluons, we can make a cut on $\tau$, and only keep those events to the left of the cut.  The fraction kept is just given by the appropriate cumulative distribution:
\begin{align}
P_q(x< \tau)&=\exp\left[
-\frac{\alpha_s}{\pi}\frac{C_F}{2}\log^2\tau
\right]\\
P_g(x< \tau)&=\exp\left[
-\frac{\alpha_s}{\pi}\frac{C_A}{2}\log^2\tau
\right] = \left[
P_q(x< \tau)
\right]^{C_A/C_F}\,.\nonumber
\end{align}
That is, the fraction of gluons kept is found by raising the fraction of quarks kept to the $C_A/C_F$ power!

We can nicely display this information in a receiver operating characteristic (ROC) plot:
\begin{center}
\includegraphics[width=7cm]{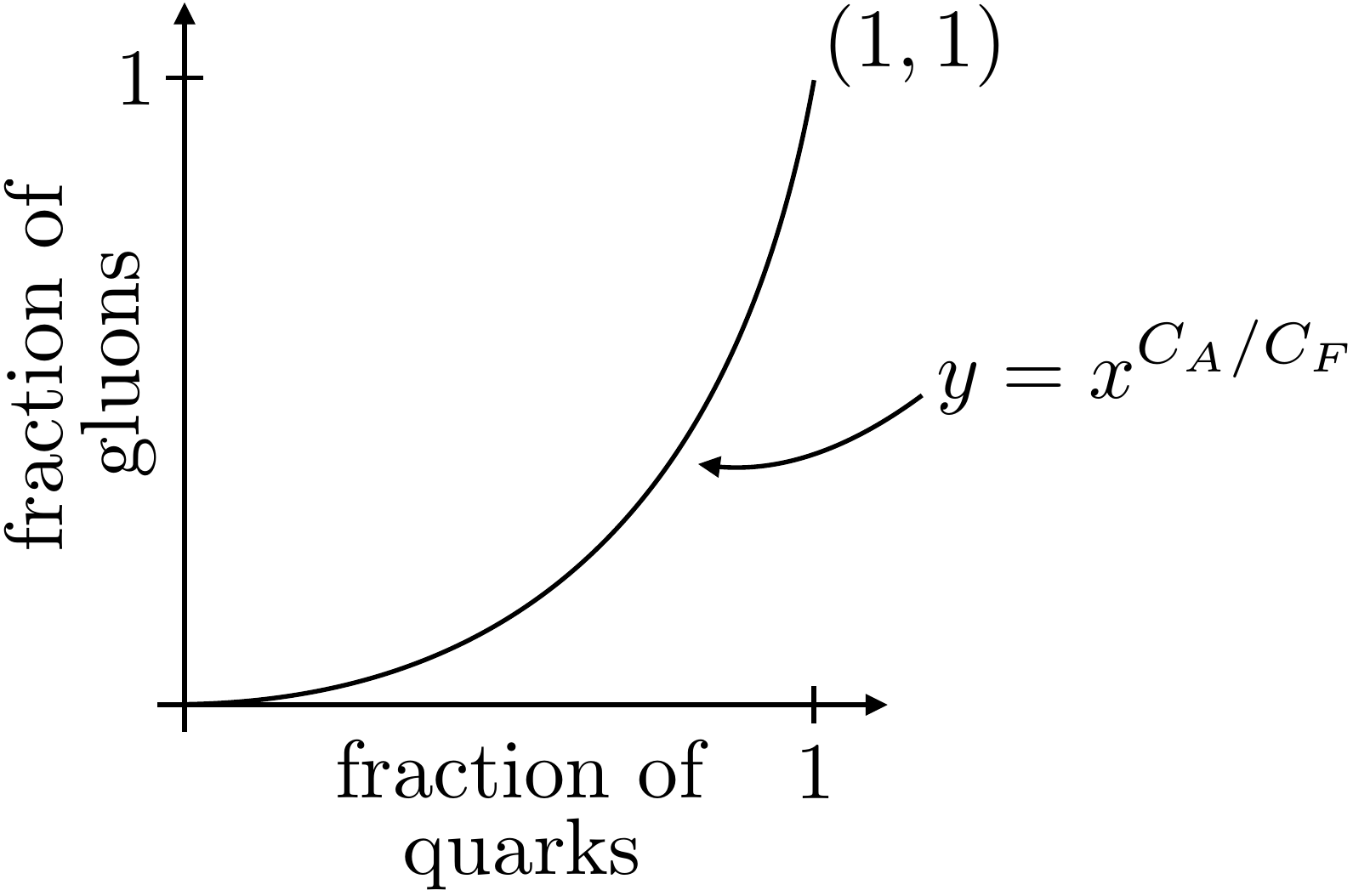}
\end{center}
This plot just displays the quark versus gluon efficiencies with this cut.  In QCD, $C_F = 4/3$ and $C_A = 3$ and so the ROC curve for quark/gluon discrimination is:
\begin{equation}
y= x^{9/4}\,.
\end{equation}
This can be improved somewhat by designing better observables or including higher-order effects, but is a benchmark for expectation.

We've gotten a lot of mileage out of our two simple axioms!

\section{Lecture 2: Jets Beyond Leading Order}

Last lecture, we developed a very simple, elegant picture of jets.  Starting from the simple assumption of scale invariance of QCD at high energies, we were able to make a robust prediction for the jet mass to all-orders in the coupling $\alpha_s$.  With one emission dominating the mass, all other emissions were forbidden to exceed the contribution from this leading emission.  This produced an exponentiation of a no-emission probability, which is called the Sudakov form factor.  At this leading order, differences between quark and gluon initiated jets arise from their different color charges; gluons have a larger color charge, and so emit more than quarks.  This leads to a larger suppression of the mass distribution from the Sudakov form factor for gluons, and a technique for discriminating quark- from gluon-initiated jets.

In this lecture, we will work to understand features of jets beyond this first approximation.  This will necessitate the introduction of jet grooming as a technique to clean up jets produced at the LHC and simplify their structure.

Let's first visualize what our simple picture of a jet is, produced at the LHC:
\begin{center}
\includegraphics[width=10cm]{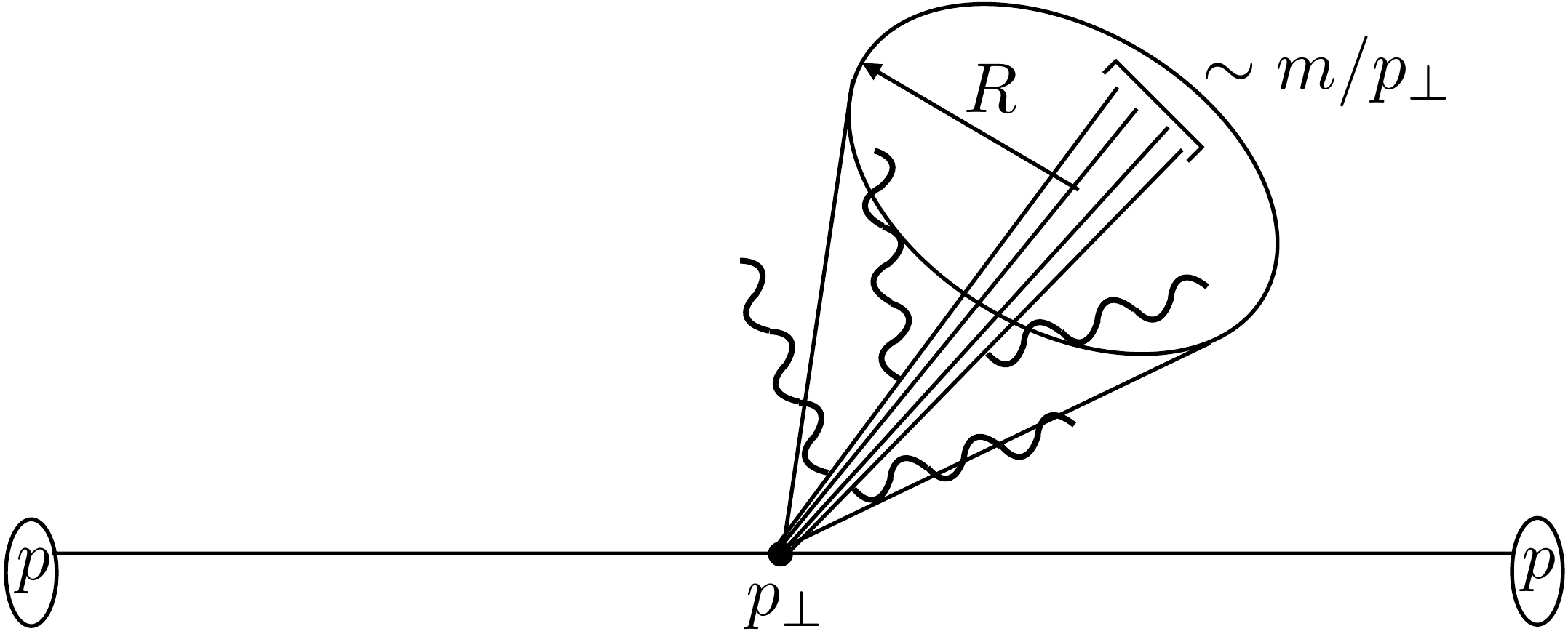}
\end{center}
A jet is some restricted region of our detector whose size is fixed by a radius $R$.  This region is identified by a large transverse momentum $p_\perp$ of a collection of collimated particles produced from individual parton scattering from proton collisions.  The angular size $\theta$ of the collimated particles is approximately the ratio of the jet's invariant mass $m$ to its $p_\perp$:
\begin{equation}
\theta \sim \frac{m}{p_\perp}\,.
\end{equation}
So, with this simple picture of a jet as a collection of final state, collimated emissions the distribution of the mass of a jet can only depend on the dimensionless quantities $m/p_\perp$ and $R$.  We saw the $m/p_\perp$ dependence in the previous lecture, while jet radius dependence $R$ only first appears at higher orders.

At the LHC, however, this simple picture is woefully incomplete.  Protons are composite particles, consisting of numerous quarks and gluons, which are the particles that ultimately interact.  Multiple parton interactions can occur in each proton collision, contributing to what is referred to as the underlying event.  Additionally, the LHC doesn't collide individual pairs of protons, but rather bunches containing billions of protons.  Many of these protons can collide in each bunch crossing, which is referred to as pile-up.  Including underlying event and pile-up, our picture of jet production muddies significantly:
\begin{center}
\includegraphics[width=10.5cm]{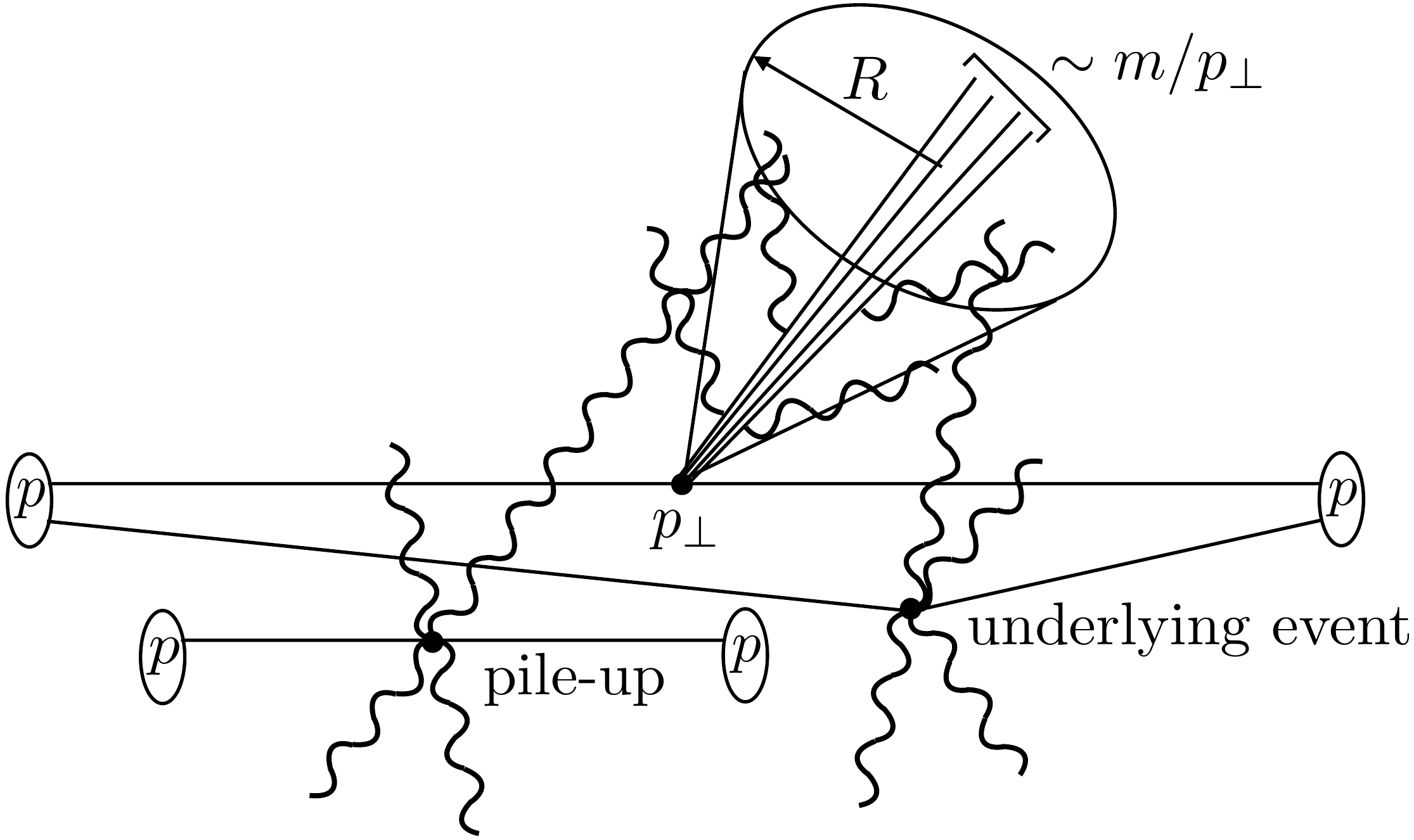}
\end{center}
Now, not only is the mass of the jet sensitive to the angular size of collinear emissions, but it is also sensitive to the energy of the underlying event and pile-up!  We can estimate this contribution to the squared jet mass $m^2$.  We can safely assume that the characteristic underlying event and pile-up transverse momentum scales are much less than the jet $p_\perp$:
\begin{equation}
p_\perp \gg p_{\perp\text{UE}},\,p_{\perp \text{pu}}\,.
\end{equation}
Also, note that the direction of the jet is (essentially) uncorrelated with the underlying event and pile-up radiation.  Therefore, on average, this radiation is just uniformly distributed of the area of the jet.  A single underlying event of pile-up emission with transverse momentum $p_{\perp\text{u/p}}$ therefore affects the jet mass as:
\begin{equation}
\Delta m^2 \simeq p_\perp p_{\perp\text{u/p}} R^2\,,
\end{equation}
or, for a transverse momentum density $\Lambda_\text{u/p}$ (transverse momentum per unit angular area) the mass is affected as
\begin{equation}
\Delta m^2 \simeq p_\perp \Lambda_\text{u/p} R^4\,.
\end{equation}

The linear mass is affected as
\begin{align}
m \to (m^2+\Delta m^2)^{1/2} &= m\left(1+\frac{\Delta m^2}{2m^2}+\cdots\right) \simeq m+\frac{\Delta m^2}{2m}\\
&\simeq m+\frac{p_\perp \Lambda_\text{u/p} R^4}{2m}\,.\nonumber
\end{align}
To get a sense of the size of this effect, take $p_\perp = 1$ TeV, $m=100$ GeV, and typically $\Lambda_\text{u/p} \sim 1$ GeV.  For a jet radius $R = 1.0$, the change to the jet mass is
\begin{equation}
\frac{\Delta m^2}{2m} \simeq \frac{1000\cdot 1}{2\cdot 100}\simeq 5\text{ GeV}\,.
\end{equation}
So, this affects the jet mass by about 5\%.  However, this is a conservative estimate.  As the luminosity of the LHC increases, the rate of pile-up collisions increases and correspondingly so does the average transverse momentum density.  Depending on the process and nominal jet mass, the effect of underlying event and pile-up could be 20\% or more, which is significant.

There's another type of radiation that can affect the jet.  There can be emissions that land in the jet that arise from re-radiation from outside the jet:
\begin{center}
\includegraphics[width=10.5cm]{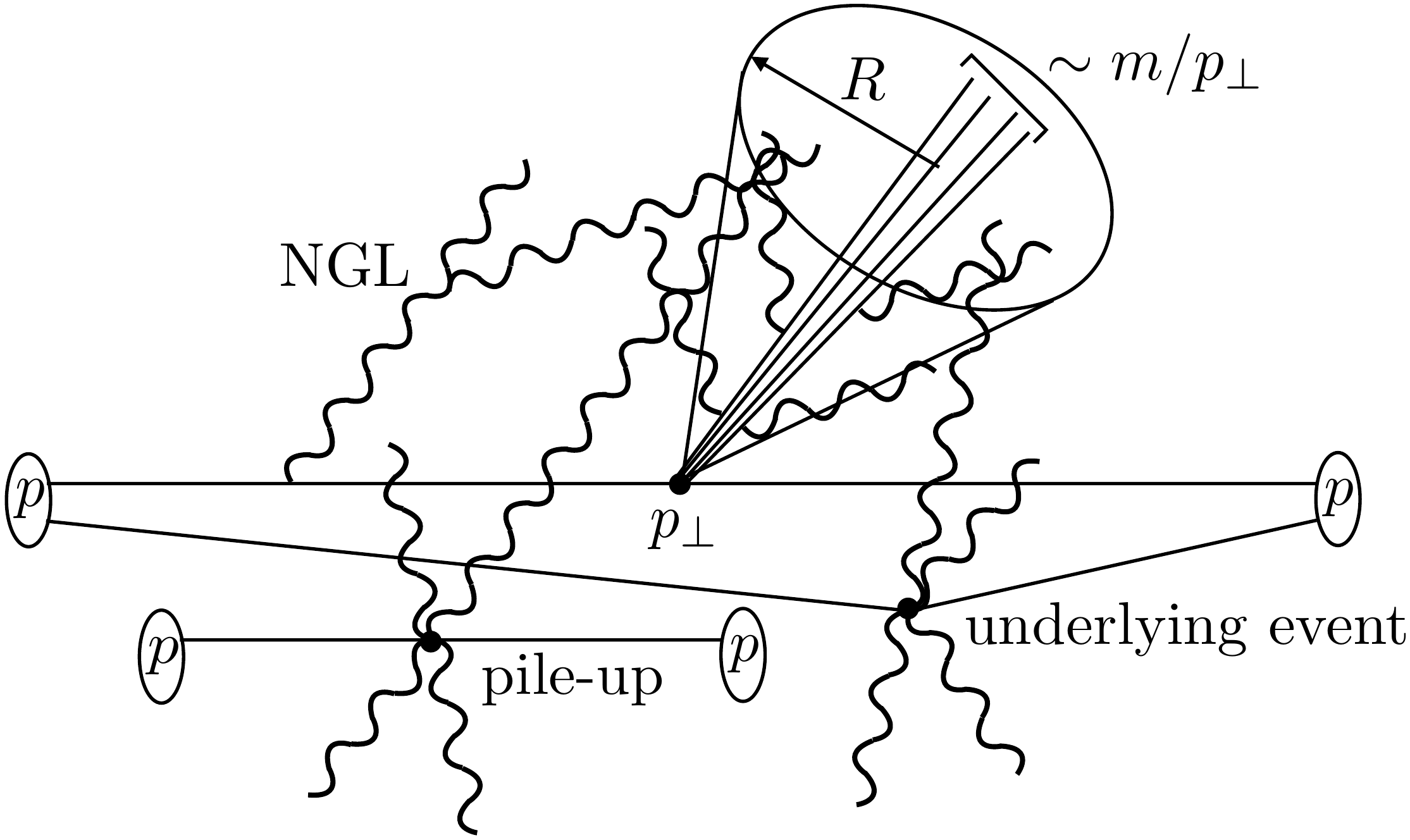}
\end{center}
The leading contribution from these re-emissions is called non-global logarithms, or NGLs \cite{Dasgupta:2001sh}.  NGLs introduce a correlation between physics in and out of the jet.  Inside the jet, scales are set by the jet mass, while out of the jet, there is no restriction on radiation.  That is, we only identify and measure the jet, so out-of-jet scales are completely unconstrained, and as high as the jet $p_\perp$ or even the scattering energy $Q$.  The leading contribution comes from two strongly-ordered gluon emissions, one of which lands in the jet.  Just like underlying event or pile-up, the direction of these NGL emissions is uncorrelated with the jet direction.  Therefore, the leading NGLs are proportional to the area of the jet:
\begin{equation}
\text{NGL}\simeq \alpha_s^2 R^2 \log^2\frac{m}{p_\perp}\,. 
\end{equation}
In general, systematically calculating NGLs is extremely non-trivial though progress has been made in understanding them recently \cite{NGLs}.  Nevertheless, along with underlying event and pile-up, NGLs present a challenge to a systematic theoretical understanding of a jet.

All of these sources of contamination have two things in common: the radiation is relatively uniformly distributed over the area of the jet and it is low energy.  This is distinct from the high energy, collinear radiation that forms the main structure of the jet (and that was our simple picture of a jet from previous lecture).  So, if we have a procedure to systematically remove soft, wide angle radiation in the jet, we can eliminate all of these sources of contamination!  Let's see how to do this.

First, let's identify wide-angle emissions in the jet.  To do this, we will recluster the particles in the jet with the Cambridge/Aachen algorithm \cite{Dokshitzer:1997in}.  That is, for all pairs $i,j$ of particles in the jet, calculate their angular separation:
\begin{equation}
d_{ij} = \frac{\Delta \eta_{ij}^2+\Delta \phi_{ij}^2}{R^2}\,,
\end{equation}
where
\begin{align}
&\Delta \eta_{ij} = \eta_i-\eta_j\,, \Delta \phi_{ij} = \phi_i-\phi_j\,.
\end{align}
Cluster the pair of particles with the smallest $d_{ij}$, that is, sum their four-vectors and replace them in the list of particles by their sum:
\begin{equation}
p_{(ij)} = p_i+p_j\,.
\end{equation}
Continue this process until all particles in the jet are clustered into the total jet four-vector.  This defines an angular ordered branching history, starting with the widest angle particles:
\begin{center}
\includegraphics[width=7cm]{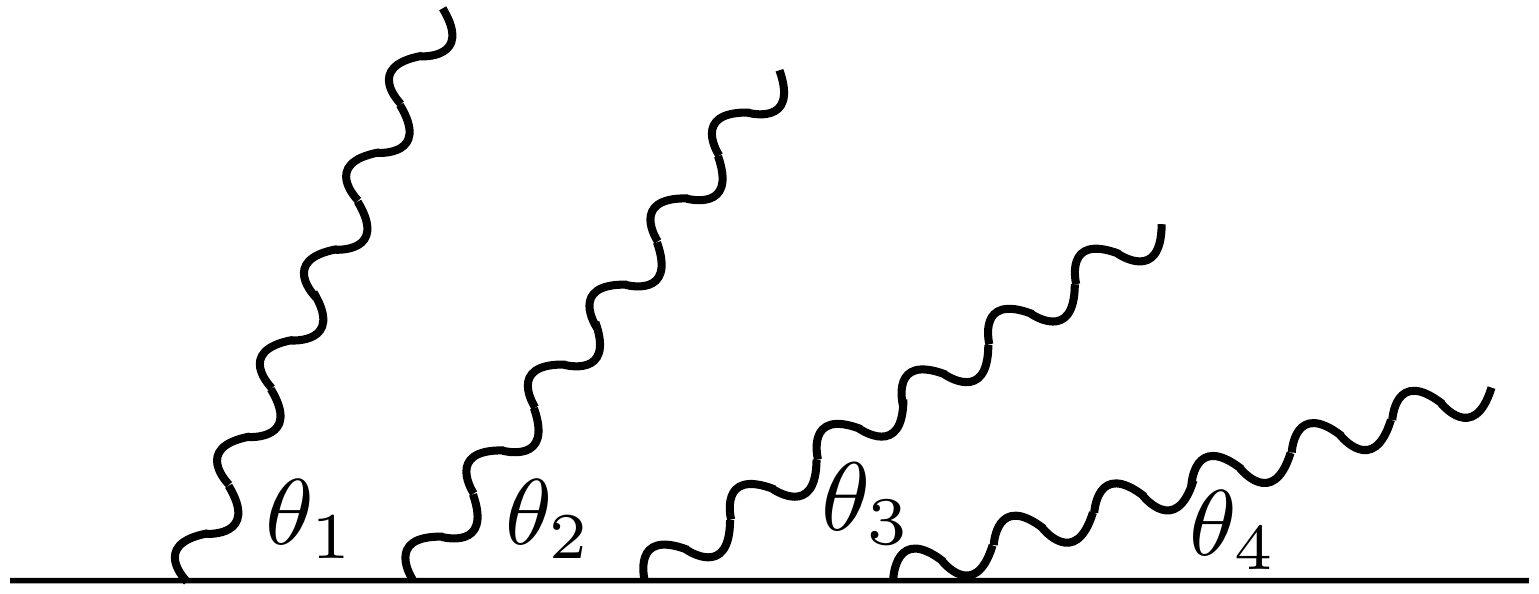} \hspace{1cm} where $\theta_1>\theta_2>\theta_3>\cdots\,.$
\end{center}
We can then systematically march through the clustering history and ask if the emission/particle is sufficiently high energy.  For a branching with particles $i,j$ we test
\begin{equation}
\frac{\min[p_{\perp i},p_{\perp j}]}{p_{\perp i}+p_{\perp j}} > z_\text{cut}(d_{ij})^{\beta/2}\,.
\end{equation}
Here, $z_\text{cut}$ and $\beta$ are parameters. $z_\text{cut}$ defines the relative energy cut, typically taken to be about 10\%.  $\beta$ defines the aggressiveness of the test.  If $\beta \to \infty$, then $(d_{ij})^{\beta/2}\to 0$, which means the test is always satisfied.  If $\beta \to 0$, then the test is just a relative energy/$p_\perp$ cut.  If the clustering fails the requirement, the softer branch is eliminated from the jet, and the procedure steps to the next smaller branch.  When a branching passes, the procedure terminates, and the remaining particles constitute the groomed jet.

This procedure is called ``soft drop grooming'' \cite{Larkoski:2014wba} and when $\beta = 0$, it is also the modifed mass drop tagger groomer (mMDT) \cite{Dasgupta:2013ihk}.  Because of the angular ordering and elimination of soft radiation, it indeed removes the radiation from underlying event, pile-up, or NGLs that contaminate the jet.  Only the hard, collinear core remains, which is just what we want.

To end this lecture, let's calculate the jet mass distribution in the presence of grooming.  We'll go back to our familiar Lund plane and identify the regions of that phase space that are no longer accessible when groomed.  For simplicity, we'll just restrict the discussion to $\beta = 0$ soft drop or mMDT grooming.

The requirement on the particles in a branching is
\begin{equation}
\frac{\min[p_{\perp i},p_{\perp j}]}{p_{\perp i}+p_{\perp j}} > z_\text{cut}\,.
\end{equation}
In the Lund plane, we restrict to the soft and collinear limit, so one of these particles carries the entire jet energy.  That is, the picture is
\begin{center}
\includegraphics[width=4.5cm]{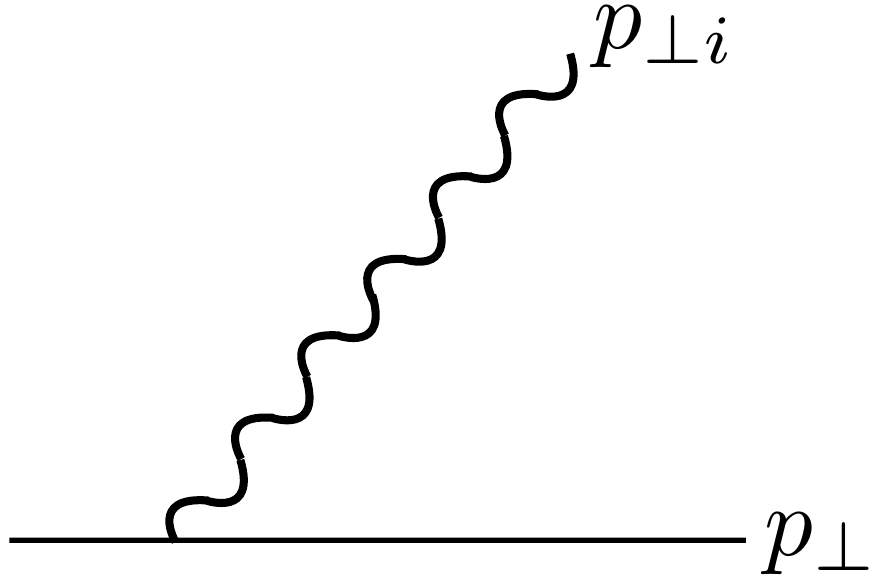}
\end{center}
and we ask if $p_{\perp i}> z_\text{cut}p_\perp$.  If this is not true, the emission is removed.  So, on the Lund plane, this eliminates a semi-infinite region from consideration:
\begin{center}
\includegraphics[width=8cm]{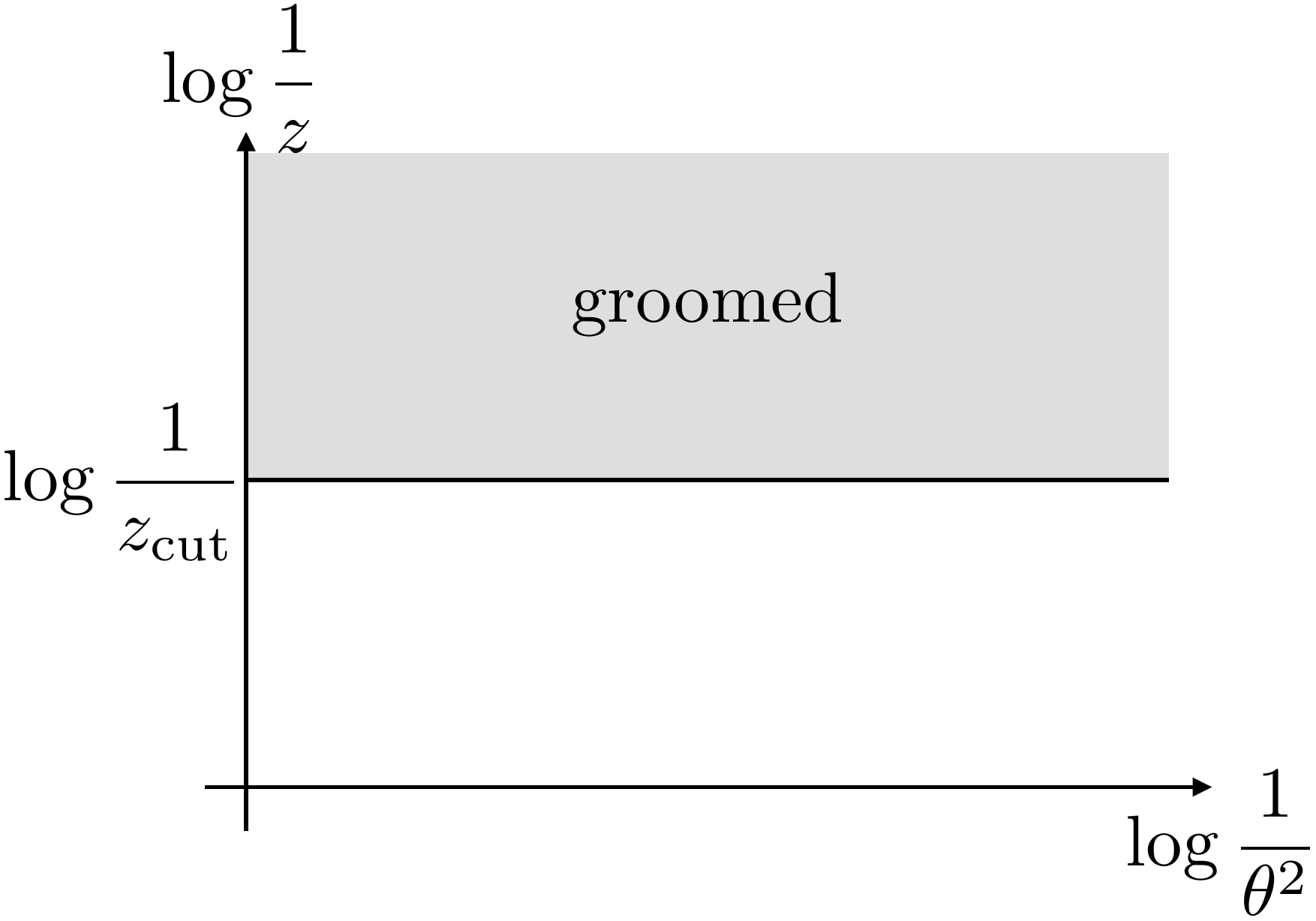}
\end{center}
Let's calculate the Sudakov factor for measuring the observable $\tau= m^2/p_\perp^2$ for this groomed phase space.

There are two distinct regions to consider.  First, if $\tau > z_\text{cut}$, we just find the Sudakov factor derived in the previous lecture:
\begin{equation}
\raisebox{-0.45\height}{\includegraphics[width=8cm]{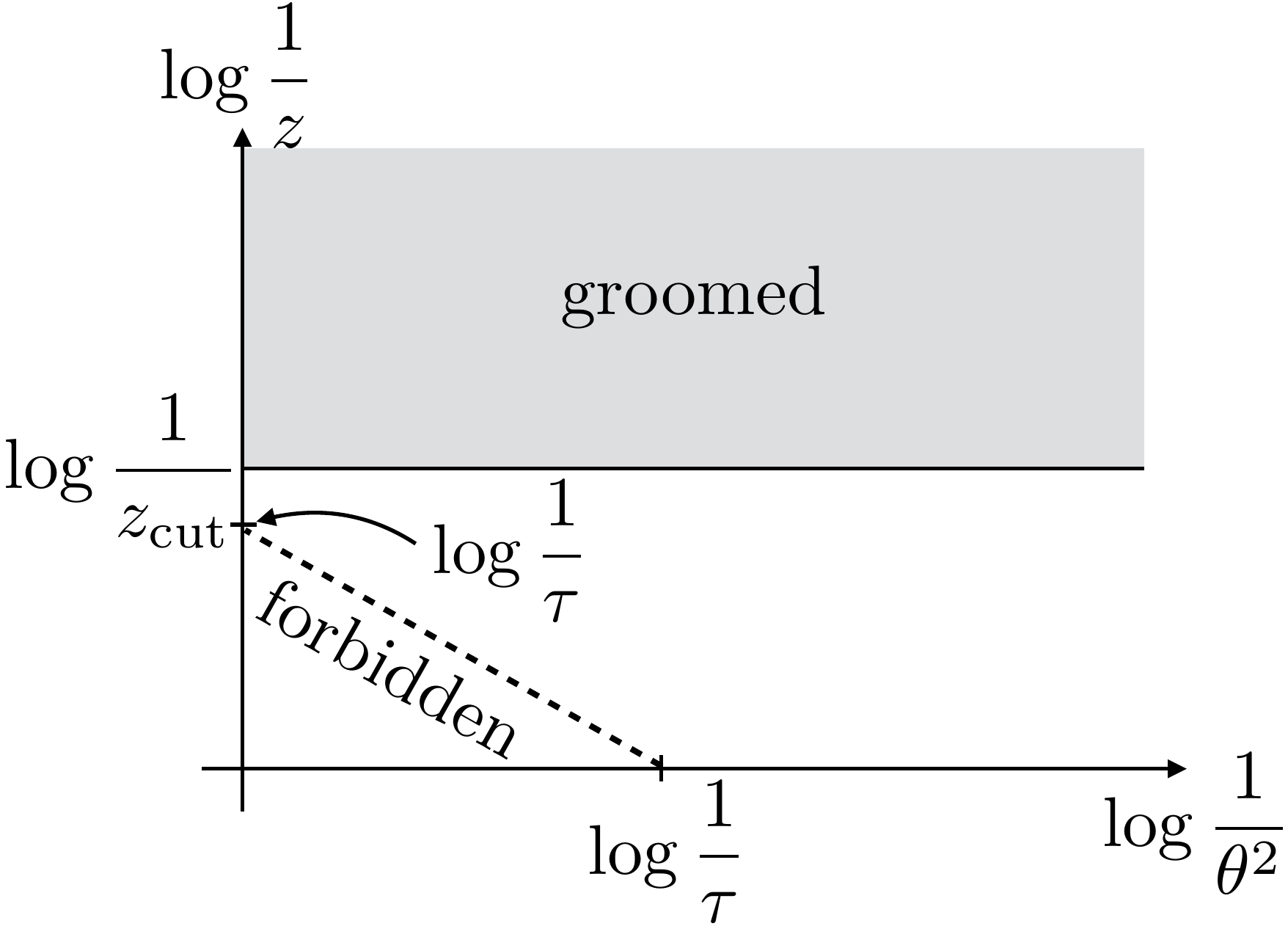}} \quad \Rightarrow \quad \Theta(\tau - z_\text{cut})\exp\left[
-\frac{\alpha_s}{\pi}\frac{C_F}{2}\log^2\tau
\right]\,.
\end{equation}
This successfully resums double logarithms of $\tau$.  However, if $\tau < z_\text{cut}$, we find something very different.  The Lund diagram is:
\begin{center}
\includegraphics[width=8cm]{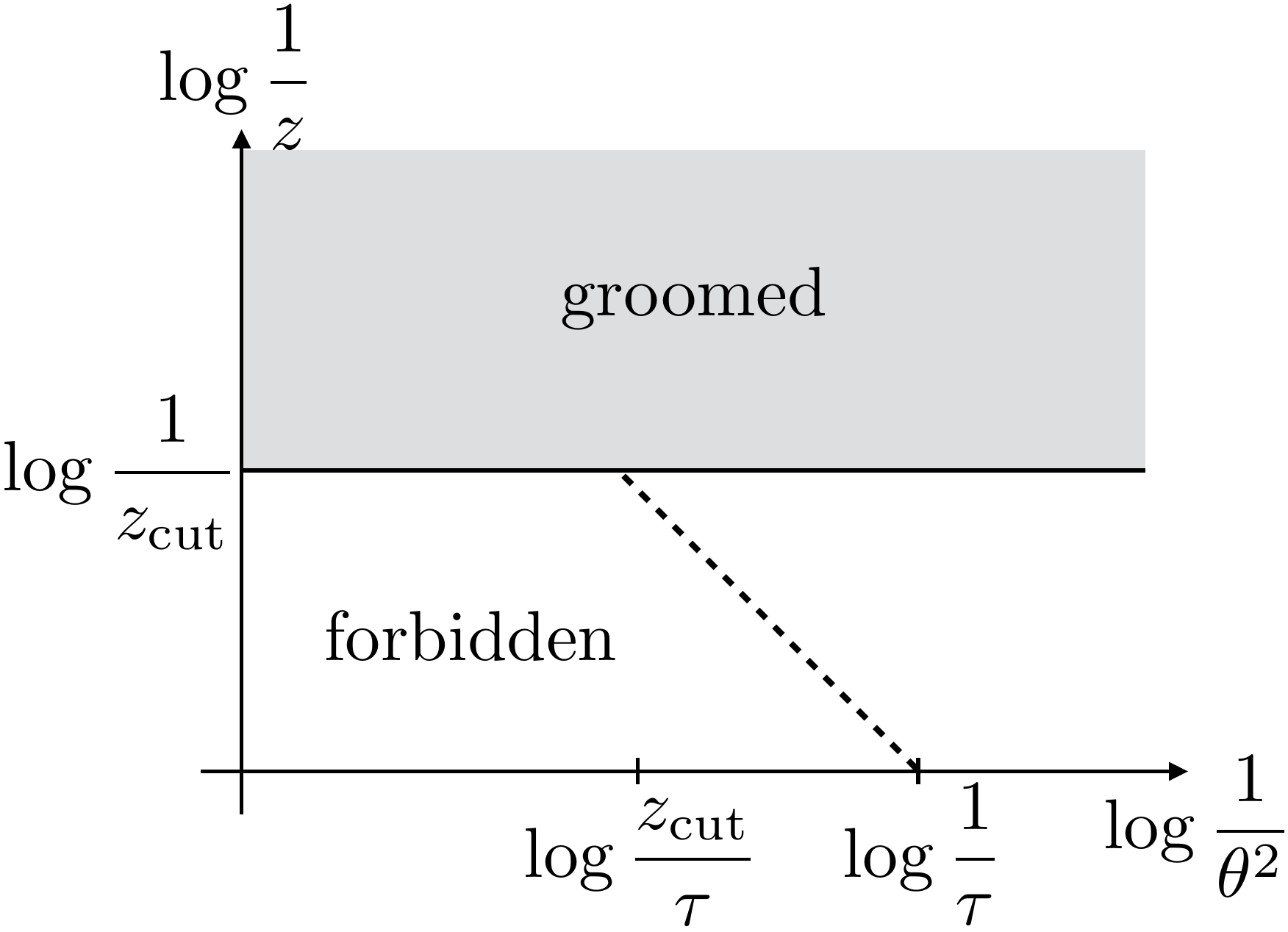}
\end{center}
Note that the angle of emission when the energy fraction $z = z_\text{cut}$ is
\begin{equation}
\tau = z_\text{cut}\theta^2 \qquad \Rightarrow \qquad \theta^2 = \frac{\tau}{z_\text{cut}}\,.
\end{equation}
The area of the forbidden region is now
\begin{equation}
\log\frac{1}{z_\text{cut}}\log\frac{z_\text{cut}}{\tau}+\frac{1}{2}\log\frac{1}{z_\text{cut}}\left(
\log\frac{1}{\tau}-\log\frac{z_\text{cut}}{\tau}
\right) = -\frac{1}{2}\log^2z_\text{cut}+\log z_\text{cut}\log \tau\,.
\end{equation}
Therefore, the Sudakov form factor in this region is
\begin{equation}
\Sigma(\tau)\supset \Theta(z_\text{cut} - \tau)\exp\left[
-\frac{\alpha_s}{\pi}C_F\left(
-\frac{1}{2}\log^2z_\text{cut}+\log z_\text{cut}\log \tau
\right)
\right]\,.
\end{equation}
There are now no double logarithms of $\tau$!  Soft drop removes soft logs of $\tau$ and replaces them with $\log z_\text{cut}$.

Putting it together, the total Sudakov factor is
\begin{align}
\Sigma(t) &= \Theta(\tau - z_\text{cut})\exp\left[
-\frac{\alpha_s}{\pi}\frac{C_F}{2}\log^2\tau
\right]\\
&
\hspace{2cm}
+\Theta(z_\text{cut} - \tau)\exp\left[
-\frac{\alpha_s}{\pi}C_F\left(
-\frac{1}{2}\log^2z_\text{cut}+\log z_\text{cut}\log \tau
\right)
\right]\nonumber\,.
\end{align}
The differential cross section is therefore
\begin{align}
\frac{d\sigma}{d\tau} = \frac{d}{d\tau}\Sigma(\tau) &= \Theta(\tau - z_\text{cut})\frac{\alpha_s C_F}{\pi}\frac{\log\frac{1}{\tau}}{\tau}\exp\left[
-\frac{\alpha_s}{\pi}\frac{C_F}{2}\log^2\tau
\right]\\
&
\hspace{1cm}
+\Theta(z_\text{cut} - \tau)\frac{\alpha_s C_F}{\pi}\frac{\log\frac{1}{z_\text{cut}}}{\tau}\exp\left[
-\frac{\alpha_s}{\pi}C_F\left(
-\frac{1}{2}\log^2z_\text{cut}+\log z_\text{cut}\log \tau
\right)
\right]\nonumber\,.
\end{align}
A plot of this distribution logarithmic in $\tau$ is
\begin{center}
\includegraphics[width=10cm]{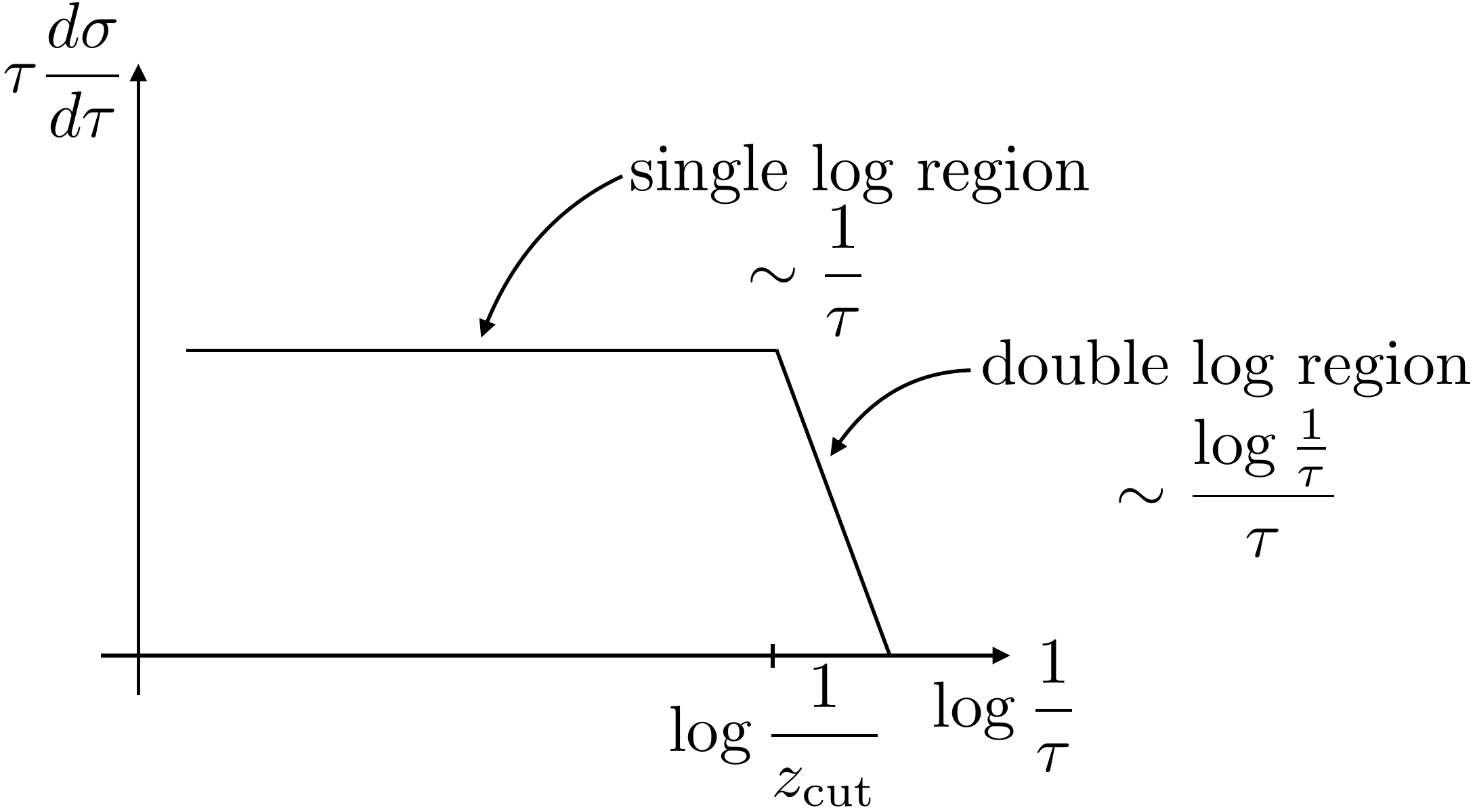}
\end{center}
Below $\tau = z_\text{cut}$, the distribution is effectively a flat line!  Thus, it is very sensitive (through its slope) to the value of $\alpha_s$, for example.  This groomer points the way to a precision jet physics program.

\end{document}